\documentstyle[psfig,editedvolume]{crckapb} 

\def\ros{{\sl ROSAT }}
\def\asca{{\sl ASCA }}
\def\arcsec{\hbox{$^{\prime\prime}$}}
\def\approxlt{\mathrel{\hbox{\rlap{\lower.55ex \hbox {$\sim$}}
        \kern-.3em \raise.4ex \hbox{$<$}}}}
\def\approxgt{\mathrel{\hbox{\rlap{\lower.55ex \hbox {$\sim$}}
        \kern-.3em \raise.4ex \hbox{$>$}}}}

\begin{opening}
\title{The nature of the luminous X-ray emission of NGC\,6240}  

\author{Stefanie Komossa}
\institute{Max-Planck-Institut f\"ur extraterrestrische Physik, Postfach 1603,
           D-85740 Garching, Germany;~~ skomossa@xray.mpe.mpg.de} 
\author{Hartmut Schulz}
\institute{Astronomisches Institut der Ruhr-Universit\"at, D-44780 Bochum,
            Germany}

\end{opening}

\runningtitle{Nature of the luminous X-ray emission of NGC\,6240}

\begin{document}


\section{Abstract}

We briefly review and extend our discussion of the \ros detection
of the extraordinarily luminous ($>$$10^{42}$\,erg/s) partly {\em extended}
($>$30\,kpc diameter) X-ray emission from the 
ultraluminous infrared galaxy NGC\,6240.
The `standard'-model of starburst outflow is contrasted with 
alternatives and a comparison with the  
X-ray properties of ellipticals is performed.

\section{Introduction}

The double-nucleus galaxy NGC\,6240
is outstanding in several respects: its infrared
H$_2$ 2.121$\mu$m and [FeII] 1.644$\mu$m line luminosities
and the ratio of H$_2$ to bolometric luminosities are the
largest currently known (van der Werf et al.\ 1993). Its huge far-infrared
luminosity of $\sim 10^{12} L_{\odot}$ (Wright et al.\ 1984) comprises
nearly all of its bolometric luminosity.
Hence, owing to its low redshift of $z$=0.024,
NGC\,6240 is one of the nearest members of the class of
ultraluminous infrared galaxies
(hereafter ULIRGs).{\footnote{We continue to refer to NGC\,6240
as ULIRG but note that, owing to the method to integrate
over the IRAS bands and the adopted value of $H_0$, most authors
now attribute an IR luminosity $<10^{12}\,L_{\odot}$  to NGC\,6240
rendering it a LIRG instead of a ULIRG}}
Its optical morphology (e.g., Zwicky et al. 1961, Fried \& Schulz 1983)
and its large stellar velocity dispersion of 360 km/s
(among the highest values ever found in the center
of a galaxy: e.g., Doyon et al. 1994) suggest that it is
a merging system on its way to become an elliptical. 
Like other ULIRGs, the object contains a compact, 
luminous CO(1-0) emitting core of molecular
gas (Solomon et al.\ 1997).
Within this core most of the ultimate
power source of the FIR radiation appears to be hidden.

There is now growing evidence that LIRGs 
are predominantly powered by star-formation 
and that the AGN contribution increases with FIR luminosity
(e.g., Shier et al. 1996, Lutz et al. 1998a, Rigopolou et al. 1998; see 
Sanders \& Mirabel 1996 and Genzel et al. 1998 for recent reviews) while
essentially all of the HyLIRGs contain QSOs (e.g., Hines et al. 1995 
and these proceedings). 
In the `transition region' around 
$L_{\rm FIR} \simeq 10^{12}\,L_{\odot}$
it then requires a careful object-by-object analysis to find out the 
majour power source.    
Concerning NGC\,6240, at least four scenarios have been suggested: 
Heating of dust by a superluminous starburst, by an AGN, 
by an old stellar population, and 
by UV radiation from molecular cloud collisions. In particular, 
previous hints for an AGN included (i) the strength of NIR recombination
lines (de Poy et al. 1986, depending on the applied reddening correction,
though), (ii) the presence of compact bright radio cores (Carral et al. 1990
but see Colbert et al. 1994), (iii) the discovery of a high-excitation core
in the southern nucleus with {\sl HST} (Barbieri et al. 1994, 1995, Rafanelli et al. 1997),
and the detection of the [OIV]\,25.9$\mu$m emission line with ISO (Lutz et al. 1996 --
but see Lutz et al. 1998b who discovered this line in a number
of starburst galaxies; Egami, this meeting).    

X-rays are a powerful tool to investigate both, the presence of an
AGN and starburst-superwind activity.   
It is the aim of the present contribution to review briefly and discuss 
our findings of evidence for a hard X-ray component in the
\ros PSPC spectrum of NGC\,6240 (Schulz et al. 1998, SKBB hereafter) 
and the detection of luminous extended
emission based on \ros HRI data (Komossa et al. 1998) in combination with
the discovery of an FeK line and hard X-ray 
component by \asca (first reported by Mitsuda 1995). 

Luminosities given below were calculated using $H_0 = 50$ km/s/Mpc.

\begin{figure}
\vspace{0.1cm} \hspace{1.5cm}
 \vbox{\psfig{figure=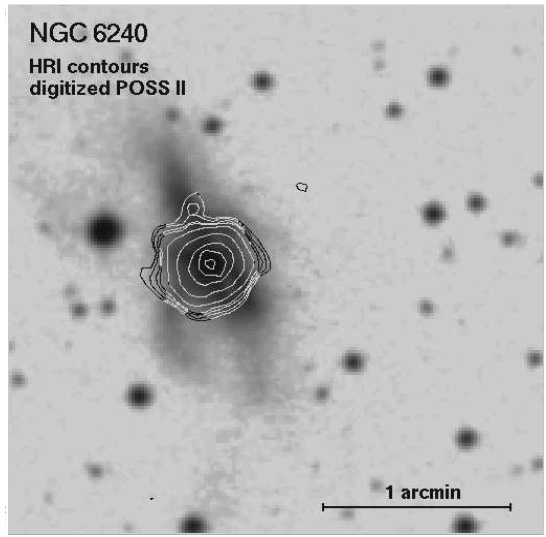,width=8.0cm,
  bbllx=1.5cm,bblly=7.9cm,bburx=17.0cm,bbury=23.2cm,clip=}}\par
\caption[]{ \ros HRI X-ray contours showing the presence of luminous
extended emission overlaid on an optical image of
NGC\,6240. The lowest contour is at $\sim$2$\sigma$ above the background. }
\label{ovl}
\end{figure}

\section{ Scenarios to explain the luminous X-ray emission
           of NGC\,6240}

\subsection {Spectral properties and origin of the hard component}
In our analysis of \ros PSPC data of NGC\,6240
we tested a large variety of models to explain the X-ray spectrum
(SKBB).
{\em One-component} fits turned out to be unlikely.
E.g., a single Raymond-Smith model requires a huge absorbing column
along the line-of-sight, the consequence being an intrinsic (absorption
corrected) luminosity of $L_{\rm X,0.1-2.4} \simeq 4\,10^{43}$ erg/s,
almost impossible to reach in any starburst-superwind scenario.
The one model that does {\em not} require excess absorption is a single
black body. Although physically implausible, this description does allow
to derive a lower limit on the intrinsically emitted X-ray luminosity
which any model has to explain:
$L_{\rm X} \approxgt 2.5\,10^{42}$ erg/s in the (0.1-2.4) keV band
(Fricke \& Papaderos 1996 
obtained 3.8\,10$^{42}$ erg/s by fitting a
thermal bremsstrahlung model and allowing for some excess absorption).
Successful {\em two-component} models require the presence
of a {\em hard} X-ray component, in form of either
 very hot thermal emission ($kT \simeq$ 7 keV) or a
 powerlaw.{\footnote{The requirement of a second component in
the \ros band can be omitted if strongly depleted metal abundances
are allowed for. This has been reported for other objects as well and
the low inferred X-ray abundances are quite puzzling given that
other methods yield much higher abundances (as also stressed by 
Netzer at this meeting). 
A recent discussion of this issue is given in Komossa \& Schulz (1998)
and Buote \& Fabian (1998) who give arguments in favour of
two-component X-ray spectral models of $\sim$solar abundances in the Raymond-Smith
component instead of single-component models of very subsolar abundances.
In particular, Buote \& Fabian conclude that 
the necessity of a second X-ray component cannot be
circumvented by details of the modelling of the Fe\,L emission.}}

Due to its large luminosity of several 10$^{42}$ erg/s we interpret
the hard component to arise from an AGN.
Both, the essential lack of non-X-ray evidence for an {\em unobscured} AGN,
and the high equivalent width of the FeK line observed by {\asca}
({\normalsize e.g., Mitsuda et al. 1995}) suggest we see the AGN mainly in scattered light.
Indeed, the X-ray spectrum can be successfully described by a `warm scatterer'
(cf. Fig. 4 of Komossa et al. 1998),
highly ionized material seen in reflection
which could also explain the strong FeK line seen by {\sl ASCA}.
Our model, which was suggested to explain the hard component,
is quite similar to the one suggested by Netzer et al. (1998)
who, however, explained the {\em whole} \asca spectrum in terms of scattering.
In this respect, we emphasize that the widely {\em extended} X-ray 
component detected with the \ros HRI (Komossa et al. 1998; cf. next Sect.) 
is likely of different origin, given the low efficiency of a hugely
extended scattering mirror (SKBB).

\begin{figure}
      \vbox{\psfig{figure=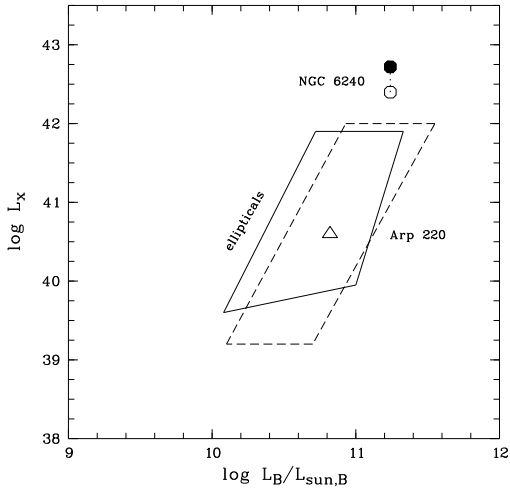,width=7cm,
          bbllx=2.9cm,bblly=1.0cm,bburx=14.8cm,bbury=12.3cm,clip=}}\par
\hfill
\begin{minipage}[]{0.40\hsize}\vspace*{-8.1cm}    
\hfill
\caption[]{ 
Position of NGC\,6240 in the $L_{\rm x}$--$L_{\rm blue}$ diagram,
compared with two samples of elliptical galaxies (solid line: Canizares
et al. 1987,
dashed: Brown \& Bregman 1998); the X-ray brightest ellipticals are those
in the
group/cluster environment. The open circle gives the minimum observed (0.1-2.4 keV)
X-ray luminosity of NGC\,6240, the filled circle the one for the 
favoured two-component model (Sect. 3.1; see SKBB, last row of their Tab. 2.)
The open triangle corresponds to Arp\,220 with $L_{\rm x}$ from Heckman et al. (1996).
$L_{\rm B}$ was calculated using $m_{\rm B,0}$ as given in NED.}
\label{lxlb}
\end{minipage}
\end{figure}

With an X-ray luminosity in scattered emission of a few 10$^{42}$ erg/s
one obtains an {\em intrinsic} luminosity of order 10$^{44-45}$ erg/s,
depending on the covering factor of the scatterer.
Various fits of \asca spectra (e.g., Mitsuda 1995, Kii et al. 1997,
Iwasawa
1998, Netzer et al. 1998, Nakagawa, these proceedings)
revealing the extension of the hard component up to 10 keV
support this conclusion; the various approaches differ in the description
of the soft component(s) and the amount of absorption of the
hard component, though.
In any case, the AGN contributes an appreciable fraction of the total
$L_{\rm bol}{\rm (NGC\,6240)} = 4\,10^{45}$ erg/s. If 
$L_{\rm bol}{\rm (AGN)} \simeq L_{\rm Edd}$ a black hole mass of
$M_{\rm bh} \simeq 10^7 M_{\odot}$ results. NGC\,6240 is expected
to form an $L_*$ elliptical galaxy rather than a giant elliptical
after having completed its merging epoch (Shier \& Fischer 1997).
However, to match the
relation $M_{\rm bh} \approx 0.002\, M_{\rm gal}$ (Lauer et al.\ 1997)
for the evolved elliptical the black hole has still
to grow by an order of magnitude
which would require another $10^9$ yrs of
accretion while the merger is settling down. Alternatively, the present
accretion rate could be below the Eddington rate. 

The inferred X-ray luminosity is  an appreciable fraction of the FIR luminosity,
suggesting that both, the starburst (e.g., Lutz et al. 1996) 
and the AGN power the FIR emission
of this ULIRG.

\subsection{Extended emission}

The HRI {\em images} (Komossa et al. 1998)
reveal that part of the huge X-ray luminosity arises in a
roughly spherical source
with strong ($\ge 2\sigma$ above background) emission out to a
radius of 20\arcsec~($\sim$14 kpc; Fig. \ref{ovl}).
Hence, NGC\,6240 is the host of one of
the {\em most luminous extended} X-ray sources
in isolated galaxies (see Fig. \ref{lxlb} where $L_{\rm X}$ is compared
with a sample of elliptical galaxies and Arp\,220).
Analytical estimates based on the Mac Low \& McCray (1988) models
show that the extended emission can be explained by superwind-shell interaction
from the central starburst (SKBB).
A puzzle is the high circular symmetry of the X-ray bubble,
in contrast to the bicone-symmetry expected in a wind-driven
supershell scenario so that it seems worthwhile to look for 
further potential contributors to the X-ray emission.  

An additional small contribution may come from a wind
induced by the large velocity dispersion of 350 km/s (Lester \& Gaffney
1994)
leading to shocks in the gas expelled by the red giant population.
Another interesting point is that the extended X-ray bubbles around
elliptical
galaxies are usually brighter in the inflow phases or `when caught
in the verge of experiencing their central cooling
catastrophe' (Ciotti et al.\ 1991, Friaca \& Terlevich 1998).
Although time scales and details for an ongoing merger are certainly
different, it is conceivable that NGC\,6240 experiences a lack of heating
when a majour starburst period has ended. In this case, a cooling flow
would commence boosting $L_{\rm x}$ and presumably shock heating the ISM
in the central kiloparsecs. Due to fragmentation, shock velocities
could be enhanced causing
the LINER like line ratios in the two nuclei (gravitational centers)
and, with lower velocities, excite the molecular cloud complex between
the nuclei
leading to the extreme H$_2$ luminosity found there (van der Werf et al.\
1993).

Its exceptional X-ray properties
make NGC\,6240 a prime target for future X-ray satellites 
like {\sl XMM} and {\sl AXAF}.

\begin{acknowledgements}
St.K. acknowledges support from the Verbundforschung under grant No.
50\,OR\,93065 and thanks the organizers for the very efficient workshop and the pleasant
workshop atmosphere.
\end{acknowledgements}

{}

\vspace*{2cm} 
\noindent    {\sl To appear in the proc. of the Ringberg workshop on `Ultraluminous Galaxies: Monsters or Babies'
              (Ringberg castle, Sept. 1998); Ap\&SS, in press}

\end{document}